\documentclass{emulateapj}

\begin{document}

\shortauthors{Luhman et al.}
\shorttitle{Brown Dwarf Disk Fractions}

\title{The Disk Fractions of Brown Dwarfs in IC 348 and 
Chamaeleon I\altaffilmark{1}}

\author{
K. L. Luhman\altaffilmark{2,3},
C. J. Lada\altaffilmark{2},
L. Hartmann\altaffilmark{2},
A. A. Muench\altaffilmark{2}, 
S. T. Megeath\altaffilmark{2}, 
L. E. Allen\altaffilmark{2},
P. C. Myers\altaffilmark{2},
J. Muzerolle\altaffilmark{4},
E. Young\altaffilmark{4},
and G. G. Fazio\altaffilmark{2}}

\altaffiltext{1}
{This work is based on observations made with the {\it Spitzer Space Telescope},
which is operated by the Jet Propulsion Laboratory, California Institute 
of Technology under NASA contract 1407.
Support for this work was provided by NASA through contract 1256790 issued
by JPL/Caltech. Support for the IRAC instrument was provided by NASA through
contract 960541 issued by JPL.}
 
\altaffiltext{2}{Harvard-Smithsonian Center for Astrophysics, 60 Garden St.,
Cambridge, MA 02138; kluhman, clada, lhartmann, gmuench, tmegeath, leallen,
pmyers, gfazio@cfa.harvard.edu.}

\altaffiltext{3}{Current address: Department of Astronomy and Astrophysics,
The Pennsylvania State University, University Park, PA 16802.}

\altaffiltext{4}{Steward Observatory, The University of Arizona, 
Tucson, AZ 85721; jamesm, eyoung@as.arizona.edu.}

\begin{abstract}

Using the Infrared Array Camera (IRAC) aboard the {\it Spitzer Space Telescope},
we have obtained mid-infrared photometry for 25 and 18 low-mass members 
of the IC~348 and Chamaeleon~I star-forming clusters, respectively
($>$M6, $M\lesssim0.08$~$M_\odot$).
We find that $42\pm13$\% and $50\pm17$\% of the two samples exhibit 
excess emission indicative of circumstellar disks.
In comparison, the disk fractions for stellar members of these clusters
are $33\pm4$\% and $45\pm7$\% 
(M0-M6, 0.7~$M_\odot\gtrsim M\gtrsim0.1$~$M_\odot$).
The similarity of the disk fractions of stars and brown dwarfs 
is consistent with a common formation mechanism and
indicates that the raw materials for planet formation are available
around brown dwarfs as often as around stars.

\end{abstract}

\keywords{accretion disks -- planetary systems: protoplanetary disks -- stars:
formation --- stars: low-mass, brown dwarfs --- stars: pre-main sequence}

\section{Introduction}
\label{sec:intro}

Because planets are born in dusty circumstellar disks, 
the likelihood of planet formation around brown dwarfs relative to that
among stars can be constrained in part by comparing the prevalence of disks 
between these two mass regimes. 
Given the central role of disks in the star formation process,
a comparison of disk fractions of brown dwarfs and stars also would 
help determine if they share common formation mechanisms. 
Extensive work has been done in measuring disk fractions for stars
\citep[e.g.,][]{kh95,hai01}, which
typically consists of infrared (IR) photometry of a significant 
fraction of a young stellar population and identification of the objects
with excess emission indicative of cool, dusty disks. 
In recent years, this method of detecting disks has been applied to
objects near and below the hydrogen burning mass limit
using photometry 
at 2-3~\micron\ \citep{luh99,luh04tau,lada00,lada04,mue01,liu03,jay03a},
4-15~\micron\ \citep{per00,com00,nat01,pas03,apa04,ster04,moh04},
and millimeter wavelengths \citep{kle03}.
However, detections of disks with the data at 2-3~\micron\
have been difficult because the emitting regions for these wavelengths
become very small for disks around low-mass bodies.
Meanwhile, disk excesses are larger at longer wavelengths, but 
have been measured for only a small number of the brighter, more massive 
objects because of technological limitations.
In comparison, because the {\it Spitzer Space Telescope} is far more sensitive
beyond 3~\micron\ than any other existing facility and can survey large areas 
of sky, it can reliably and efficiently detect disks for brown dwarfs at very 
low masses \citep{luh05ots} and for large numbers of brown dwarfs in young 
clusters.

To capitalize on the unique capabilities of {\it Spitzer} for measuring
disk fractions, we have used the 
Infrared Array Camera \citep[IRAC;][]{faz04} to obtain mid-IR photometry 
for spectroscopically confirmed stellar and substellar members of the 
star-forming clusters IC~348 
\citep[e.g.,][]{her98,luh03}
and Chamaeleon~I 
\citep[e.g.,][]{com04,luh04cha}.
In this Letter, we describe these observations, identify the cluster members 
that exhibit mid-IR excesses indicative of dusty inner disks, compare
the disk fractions of brown dwarfs and stars, and discuss the resulting
implications for the formation mechanism of brown dwarfs and 
planet formation around brown dwarfs.

\section{Observations}
\label{sec:obs}

As a part of the Guaranteed Time Observations of the IRAC instrument team,
we obtained images of IC~348 and Chamaeleon~I at 
3.6, 4.5, 5.8, and 8.0~\micron\ with IRAC on the {\it Spitzer Space Telescope}.
We performed seven sets of observations: three large shallow maps 
of IC~348 and the northern and southern clusters in Chamaeleon~I, one small
deep map of IC~348, two small deep maps of the southern cluster in 
Chamaeleon~I, and a single position toward the low-mass binary 
2MASS~J11011926-7732383
on the southwestern edge of Chamaeleon~I \citep{luh04bin}.
The characteristics of these maps are summarized in Table~\ref{tab:log}.
Further details of the observations and data reduction for IC~348 and the 
northern cluster of Chamaeleon~I are provided by Lada et al.\ (in preparation) 
and \citet{luh05ots}, respectively. Similar methods were used for 
the remaining maps in Table~\ref{tab:log}.
For all data, we have adopted zero point magnitudes ($ZP$)
of 19.670, 18.921, 16.855, and 17.394 in the 3.6, 4.5, 5.8 and 8~\micron\ bands,
where $M=-2.5 \log (DN/sec) + ZP$ \citep{rea05}. These values of $ZP$ 
differ slightly from those used for OTS~44 by \citet{luh05ots} and 
for Taurus by \citet{har05}. In Tables~\ref{tab:ic348} and \ref{tab:cha}, 
we list IRAC photometry for all known members of IC~348 and Chamaeleon~I 
that are likely to be brown dwarfs 
($>$M6)\footnote{The hydrogen burning mass limit
at ages of 0.5-3~Myr corresponds to a spectral type of $\sim$M6.25
according to the models of \citet{bar98} and \citet{cha00} and the
temperature scale of \citet{luh03}.}
and that are within our images.
Measurements for the earlier, stellar members of these clusters will be 
tabulated in forthcoming studies. 
An absent measurement in Table~\ref{tab:ic348} indicates that the object
was below the detection limit in that filter. 
Because of the weaker background emission in Chamaeleon~I, the
detection limits are much better in that cluster, and all objects 
in Table~\ref{tab:cha} have extrapolated photospheric fluxes above
the detection limits for all four bands.
Thus, an absent measurement in Table~\ref{tab:cha} indicates
contamination by cosmic rays or a position beyond the map's field of view.

\section{Analysis}

To measure disk fractions in IC~348 and Chamaeleon~I, we first define 
the samples of stars and brown dwarfs that will be used. 
We consider all known members of IC~348 
\citep[][references therein]{luh03,luh05flam} and 
Chamaeleon~I \citep[][references therein]{luh04cha,luh04ots,com04}
that have measured spectral types of M0 or later ($M\lesssim0.7$~$M_\odot$)
and detections in our IRAC images. This spectral type range encompasses 
most of the known members of each cluster ($>80$\%).
Because many of the known members of Chamaeleon~I were 
originally discovered through the presence of signatures 
directly or indirectly related to disks (IR excess, H$\alpha$ emission), 
membership samples from earlier studies of the cluster are potentially biased 
toward objects with disks, which would preclude a meaningful disk fraction
measurement. Therefore, to ensure that we have an unbiased sample of members
of Chamaeleon~I, we include in our analysis the additional members 
discovered during a new magnitude-limited survey of the cluster
(Luhman, in preparation)\footnote{In this survey, candidate low-mass stars
and brown dwarfs across all of Chamaeleon~I were identified through 
color-magnitude diagrams constructed from $JHK_s$ photometry from the 
Two-Micron All-Sky Survey (2MASS) and $i$ photometry from the Deep 
Near-Infrared Survey of the Southern Sky (DENIS). Additional candidates
at fainter levels were identified with deeper optical and near-IR images
of smaller fields toward the northern and southern clusters. 
These candidates were then classified as field stars or members through
followup spectroscopy. The resulting completeness was similar to that
achieved on other surveys using the same methods \citep[e.g.,][]{luh03,ls04}.}.
Finally, for the purposes of this work, we
treat as members of IC~348 the two candidates from \citet{luh05flam}, sources 
1050 and 2103. The resulting samples for IC~348 and Chamaeleon~I contain
246 and 109 objects, respectively.

\begin{figure}
\plotone{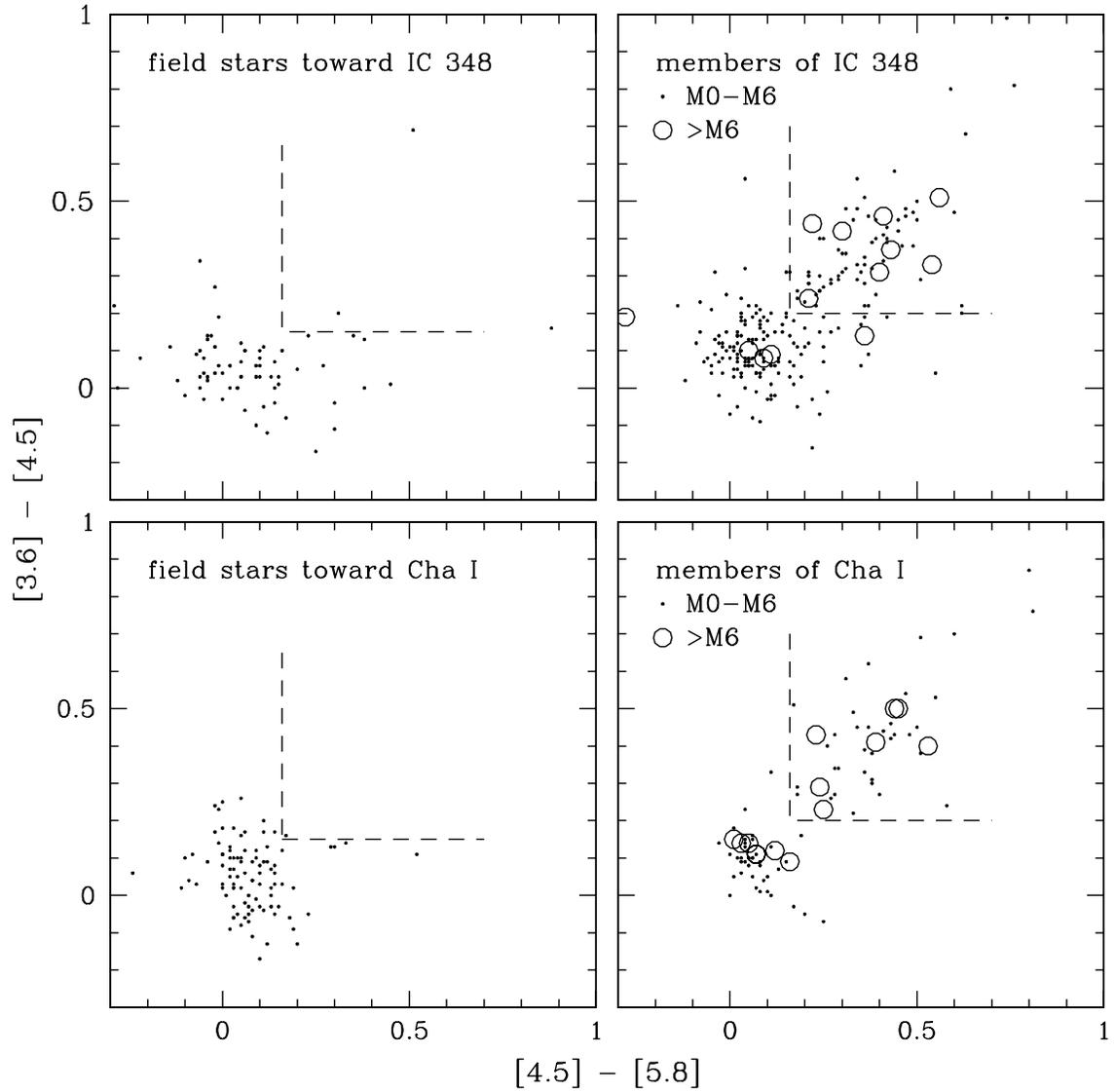}
\caption{ 
{\it Spitzer} IRAC color-color diagrams for the IC~348 and Chamaeleon~I 
star-forming clusters. 
{\it Left:} Known field stars toward each cluster illustrate the spread 
in colors that arises from photometric errors.
Relative to average colors of $[3.6]-[4.5]=0.05$ and $[4.5]-[5.8]=0.06$, 
nearly all of the field stars have color excesses less than 0.1 in at least 
one color ({\it dashed line}).
{\it Right:}
Among the known members of each cluster, objects with disks are identified by
excesses greater than 0.1 in both colors, which correspond to 
$[3.6]-[4.5]\gtrsim0.2$ and $[4.5]-[5.8]\gtrsim0.16$ for the M spectral types
in these samples ({\it dashed line}). 
}
\label{fig:quad}
\end{figure}

To identify objects with disks in the samples we have defined for IC~348
and Chamaeleon~I, we use an IRAC color-color diagram consisting of 
[3.6]-[4.5] versus [4.5]-[5.8].
The dependence of these colors on extinction and spectral type 
is very small for the range of extinctions and types in question.
Most of the objects in our samples exhibit $A_V<4$, which
corresponds to $E([3.6]-[4.5])<0.04$ and $E([4.5]-[5.8])<0.02$.
The effect of spectral type on these colors was determined by computing
the average colors as a function of spectral type of objects within
the (diskless) clump of members near the origin in the color-color
diagram of each cluster in Figure~\ref{fig:quad}. This analysis indicates that
the intrinsic [3.6]-[4.5] can be fit by two linear relations between 
$[3.6]-[4.5]=0.01$, 0.105, and 0.13 at M0, M4, and M8, respectively, while
the [4.5]-[5.8] colors show no dependence on spectral type and
have an average value of 0.06.
In comparison, colors using bands shorter than [3.6] are more sensitive to 
extinction and spectral type, and thus
are less attractive choices for this analysis.
Meanwhile, measurements at 8.0~\micron\ are available for fewer objects 
than the three shorter IRAC bands, primarily because of the bright reflection
nebulosity in IC~348.

In Figure~\ref{fig:quad}, we plot [3.6]-[4.5] versus [4.5]-[5.8] for 
the samples in IC~348 and Chamaeleon~I. One of the two colors is unavailable
for 24 and 18 objects (12 and 4 at $>$M6) in these samples, respectively, and 
thus are not shown. 
In addition to these samples of cluster members, we include in 
Figure~\ref{fig:quad} all objects in the IRAC images that have been classified
as field stars in previous studies \cite[e.g.,][]{luh03,luh04cha}, which 
correspond to 81 and 99 stars toward IC~348 and Chamaeleon~I, respectively.
We use these field stars as diskless control samples to gauge the 
scatter in colors due to photometric errors.
The scatter in the field stars toward Chamaeleon~I is actually larger than
that of the clump of members near the origin, probably because the 
field stars are more heavily weighted toward fainter levels.
According to the distributions of [3.6]-[4.5] and [4.5]-[5.8] for the
fields stars in Figure~\ref{fig:quad}, excesses greater than 0.1 in both
colors represent a significant detection of disk emission.
These color excesses also coincide with a natural break in the 
distribution of colors for members of Taurus \citep{har05} and Chamaeleon~I 
(Figure~\ref{fig:quad}). A break of this kind is present but less well-defined 
in IC~348, probably because of the larger photometric errors caused by the 
brighter background emission.
Therefore, we used these color excess criteria to identify objects
with disks among the members of Chamaeleon~I and IC~348 in 
Figure~\ref{fig:quad}. To compute the color excesses of the cluster members,
we adopted the intrinsic colors as a function of spectral type derived earlier 
in this section.  This analysis produces disk fractions of 69/209 (M0-M6) and 
8/13 ($>$M6) in IC~348 and 35/77 (M0-M6) and 7/14 ($>$M6) in Chamaeleon~I.
Among the members that are not plotted in Figure~\ref{fig:quad}
(i.e., lack photometry at 3.6, 4.5, or 5.8~\micron),
4/11 and 2/11 in IC~348 and 6/14 and 2/4 in Chamaeleon~I exhibit significant 
excesses in the other available colors (e.g., $K$-[4.5], [4.5]-[8.0]), 
while sources 621 and 761 in 
IC~348 have uncertain IRAC measurements and therefore are excluded.
After accounting for these additional objects, we arrive at disk fractions of 
73/220=$33\pm4$\% (M0-M6) and 10/24=$42\pm13$\% ($>$M6) in IC~348 and 
41/91=$45\pm7$\% (M0-M6) and 9/18=$50\pm17$\% ($>$M6) in Chamaeleon~I.

Disks with inner holes extending out to $\sim1$~AU can be undetected 
in the colors we have used, but they can have strong excess emission
at longer wavelengths \citep{cal02,for04}. For instance, source 316 in IC~348 
exhibits excess emission at 8~\micron\ but not in the shorter bands. 
Thus, our measurements apply only to inner disks that are capable 
of producing significant excesses shortward of 6~\micron\ and represent
lower limits to the total disk fractions.

\section{Discussion}

We examine the implications of our brown dwarf disk fractions
by first considering previous measurements of this kind 
in IC~348 and Chamaeleon~I. 
Using $JHKL\arcmin$ photometry, \citet{jay03a} searched for evidence of disks
among 53 objects in IC~348, Taurus, $\sigma$~Ori, Chamaeleon~I, 
the TW Hya association, Upper Scorpius, and Ophiuchus, 27 of which are later 
than M6\footnote{For objects in Chamaeleon~I, we adopt the spectral types of 
\citet{luh04cha}.}.
When sources at all spectral types were combined (i.e., both low-mass stars
and brown dwarfs), the resulting disk fractions for individual clusters 
exhibited large statistical errors of $\sim25$\%. 
Better statistics were possible for a sample combining
Chamaeleon~I, IC~348, Taurus, and U~Sco, for which \citet{jay03a} found a 
disk fraction of 40-60\%. 
For the objects with types of $\leq$M6, we find that the disk/no disk 
classifications of \citet{jay03a} agree well with those based on our IRAC data. 
However, we find no excess emission in the IRAC colors for 2/3 objects 
later than M6 in IC~348 and Chamaeleon~I (Cha~H$\alpha$~7 and 12)
that were reported to have disks by \citet{jay03a}.

An $L\arcmin$-band survey similar to that of \citet{jay03a} was performed
by \citet{liu03}. They considered a sample of 7 and 32 late-type members of 
Taurus and IC~348, respectively, 12 of which have optical spectral types 
later than M6 \citep{bri02,her98,luh99,luh03}. 
For their entire sample of low-mass stars and brown dwarfs, \citet{liu03} 
found a disk fraction of $77\pm15$\%, which is a factor of two larger than
our measurements for IC~348. Among the 28 members of IC~348 from the sample of 
\citet{liu03} for which [3.6]-[4.5] and [4.5]-[5.8] are available, only 11
objects show excesses in these colors.
We find that 9/10 objects with $E(K-L\arcmin)>0.2$
in the data from \citet{liu03} do indeed exhibit significant excesses in the 
IRAC colors. However, the putative detections of disks with smaller $L\arcmin$ 
excesses from \citet{liu03} are not confirmed by the IRAC measurements.
Because the color excess produced by a disk grows with increasing wavelengths,
any bona fide detection of a disk at $L\arcmin$ would be easily verified
in the IRAC data.

Our IRAC images of IC~348 and Chamaeleon~I have produced 
the most accurate, statistically significant measurements to date of disk 
fractions for brown dwarfs ($>$M6). 
For both clusters, these measurements are consistent with the disk fractions
exhibited by the stellar populations 
(M0-M6, 0.7~$M_\odot\gtrsim M\gtrsim0.1$~$M_\odot$).
These results support the notion that stars and brown dwarfs
share a common formation history, but do not completely exclude
some scenarios in which brown dwarfs form through a distinct mechanism
\citep{bat03}. 
The similarity of the disk fractions of stars and brown dwarfs also 
indicates that the building blocks of planets are available
around brown dwarfs as often as around stars. 
The relative ease with which planets arise from these building blocks
around stars and brown dwarfs remains unknown.




\acknowledgements
K. L. was supported by grant NAG5-11627 from the NASA Long-Term Space
Astrophysics program.



\begin{deluxetable}{lcccccl}
\tabletypesize{\scriptsize}
\tablewidth{0pt}
\tablecaption{Summary of IRAC Maps in IC~348 and Chamaeleon~I\label{tab:log}}
\tablehead{
\colhead{Target} &
\colhead{3.6/5.8 Center} &
\colhead{4.5/8.0 Center} &
\colhead{Dimensions} &
\colhead{Angle\tablenotemark{a}} &
\colhead{Exp Time\tablenotemark{b}} &
\colhead{Date} \\
\colhead{} &
\colhead{} &
\colhead{} &
\colhead{(arcmin)} &
\colhead{(degrees)} &
\colhead{(sec)} &
\colhead{(UT)} 
} 
\startdata
IC~348 & 03 43 24 +32 07 03 & 03 44 18 +32 13 47 & $33\times29$ & 170 & 20.8 & 2004 Feb 11 \\
IC~348 & 03 44 44 +32 05 50 & 03 44 37 +32 12 27 & $15\times15$ & 79 & 1548.8 & 2004 Feb 18 \\
Cha~I-N & 11 09 26 $-$76 36 26 & 11 10 16 $-$76 30 20 & $33\times29$ & 28 & 20.8 & 2004 Jul 4 \\
Cha~I-S & 11 07 44 $-77$ 34 46 & 11 07 48 $-$77 28 03 & $33\times29$ & 3 & 20.8 & 2004 Jun 10 \\
Cha~I-S & 11 08 46 $-$77 37 19 & 11 06 45 $-77$ 39 37 & $20\times15$ & 72 & 968 & 2004 Feb 19 \\
Cha~I-S & 11 06 41 $-77$ 39 09 & 11 08 47 $-77$ 37 55 & $20\times15$ & 80 & 968 & 2004 Sep 2 \\
2M~J1101-7732A+B & 11 02 00 $-$77 35 17 & 11 00 43 $-$77 29 58 & $12\times5$ & 144 & 52 & 2005 May 9 \\
\enddata
\tablenotetext{a}{Position angle of the long axis of the maps.}
\tablenotetext{b}{Total exposure time for each position and filter.}
\end{deluxetable}


\begin{deluxetable}{llllll}
\tablewidth{0pt}
\tablecaption{IRAC Photometry for Late-Type Members of IC~348\label{tab:ic348}}
\tablehead{
\colhead{ID} &
\colhead{Sp Type\tablenotemark{a}} &
\colhead{$[3.6]$} &
\colhead{$[4.5]$} &
\colhead{$[5.8]$} &
\colhead{$[8.0]$}
} 
\startdata
291 &      M7.25 & 11.94$\pm$0.02 & 11.57$\pm$0.02 & 11.14$\pm$0.04 & 10.39$\pm$0.07   \\
316 &       M6.5 & 12.60$\pm$0.02 & 12.51$\pm$0.02 & 12.40$\pm$0.05 & 11.91$\pm$0.14   \\
329 &       M7.5 & 12.87$\pm$0.02 & 12.77$\pm$0.02 & 12.72$\pm$0.06 & 12.67$\pm$0.25   \\
355 &         M8 & 12.97$\pm$0.02 & 12.83$\pm$0.02 & 12.47$\pm$0.12 &    \nodata   \\
363 &         M8 & 13.19$\pm$0.02 & 13.11$\pm$0.02 & 13.02$\pm$0.03 & 13.02$\pm$0.13   \\
405 &         M8 & 13.43$\pm$0.02 & 13.29$\pm$0.02 &    \nodata &    \nodata   \\
407 &         M7 & 14.03$\pm$0.02 & 13.70$\pm$0.04 & 13.16$\pm$0.04 & 12.39$\pm$0.08   \\
415 &       M6.5 & 12.83$\pm$0.03 & 12.37$\pm$0.03 & 11.96$\pm$0.11 &    \nodata   \\
437 &      M7.25 & 13.61$\pm$0.02 & 13.42$\pm$0.03 & 13.70$\pm$0.08 &    \nodata   \\
468 &      M8.25 & 13.31$\pm$0.02 & 12.80$\pm$0.03 & 12.24$\pm$0.04 & 11.47$\pm$0.09   \\
478 &      M6.25 & 13.90$\pm$0.02 & 13.59$\pm$0.02 & 13.19$\pm$0.13 &    \nodata   \\
603 &       M8.5 & 14.18$\pm$0.05 & 13.94$\pm$0.07 & 13.73$\pm$0.08 &    \nodata   \\
611 &         M8 & 14.47$\pm$0.08 & 14.31$\pm$0.08 &    \nodata &    \nodata   \\
613 &      M8.25 & 14.96$\pm$0.08 & 15.10$\pm$0.19 &    \nodata &    \nodata   \\
624 &         M9 & 15.94$\pm$0.19 & 15.69$\pm$0.21 &    \nodata &    \nodata   \\
690 &      M8.75 & 14.65$\pm$0.04 & 14.23$\pm$0.05 & 13.93$\pm$0.12 &    \nodata   \\
703 &         M8 & 14.36$\pm$0.03 & 13.92$\pm$0.03 & 13.70$\pm$0.09 &    \nodata   \\
705 &         M9 & 15.13$\pm$0.06 & 14.86$\pm$0.05 &    \nodata &    \nodata   \\
738 &      M8.75 & 15.24$\pm$0.04 & 14.89$\pm$0.03 &    \nodata &    \nodata   \\
761 &         M7 & 14.73$\pm$0.10 & 14.38$\pm$0.10 &    \nodata &    \nodata   \\
906 &      M8.25 & 15.35$\pm$0.03 & 15.19$\pm$0.03 &    \nodata &    \nodata   \\
935 &      M8.25 & 14.55$\pm$0.02 & 14.54$\pm$0.03 &    \nodata &    \nodata   \\
1050\tablenotemark{b} &      $>$M8.5 & 15.07$\pm$0.07 & 14.91$\pm$0.04 &    \nodata &    \nodata   \\
2103\tablenotemark{b} &      $>$M8.5 & 15.67$\pm$0.05 & 15.58$\pm$0.04 &    \nodata &    \nodata   \\
4044 &         M9 & 15.23$\pm$0.09 & 14.79$\pm$0.04 &    \nodata &    \nodata   \\
\enddata
\tablenotetext{a}{From \citet{luh98}, \citet{luh99}, \citet{luh03}, 
and \citet{luh05flam}.}
\tablenotetext{b}{Candidate members \citep{luh05flam}.}
\end{deluxetable}


\begin{deluxetable}{llllll}
\tabletypesize{\small}
\tablecaption{IRAC Photometry for Late-Type Members of Chamaeleon~I\label{tab:cha}}
\tablehead{
\colhead{ID} &
\colhead{Sp Type\tablenotemark{a}} &
\colhead{$[3.6]$} &
\colhead{$[4.5]$} &
\colhead{$[5.8]$} &
\colhead{$[8.0]$}
} 
\startdata
  2MASS~11011926-7732383A+B &      M7.25+M8.25 & 11.00$\pm$0.01 & 10.86$\pm$0.01 & 10.83$\pm$0.01 & 10.76$\pm$0.02   \\
  CHSM~17173 &         M8 & 11.96$\pm$0.01 & 11.87$\pm$0.01 & 11.71$\pm$0.03 & 11.76$\pm$0.04   \\
  2MASS~11085176-7632502 &      M7.25 & 12.43$\pm$0.01 & 12.28$\pm$0.01 & 12.27$\pm$0.05 & 12.21$\pm$0.07   \\
     OTS~44 &       M9.5 & 13.71$\pm$0.02 & 13.21$\pm$0.02 & 12.76$\pm$0.04 & 12.04$\pm$0.02   \\
  2MASS~11084952-7638443 &         M9 & 13.61$\pm$0.01 & 13.21$\pm$0.02 & 12.68$\pm$0.06 & 12.14$\pm$0.04   \\
    ISO~217 &      M6.25 & 10.79$\pm$0.01 & 10.29$\pm$0.01 &  9.85$\pm$0.01 &  9.17$\pm$0.03   \\
 2MASS~11100658-7642486 &      M9.25 & 14.48$\pm$0.03 & 14.25$\pm$0.03 &    \nodata & 13.98$\pm$0.14   \\
 2MASS~11114533-7636505 &         M8 & 13.35$\pm$0.01 & 12.92$\pm$0.02 & 12.69$\pm$0.06 & 12.16$\pm$0.05   \\
 2MASS~11104006-7630547 &      M7.25 & 12.83$\pm$0.01 & 12.71$\pm$0.01 & 12.59$\pm$0.05 & 12.63$\pm$0.06   \\
  Cha~H$\alpha$~11 &      M7.25 & 12.98$\pm$0.01 & 12.75$\pm$0.02 & 12.50$\pm$0.04 & 12.05$\pm$0.06   \\
 2MASS~11112249-7745427 &      M8.25 & 13.55$\pm$0.01 &    \nodata & 12.80$\pm$0.04 &    \nodata   \\
  Cha~H$\alpha$~12 &       M6.5 & 11.34$\pm$0.01 & 11.27$\pm$0.03 &    \nodata & 11.13$\pm$0.03   \\
  Cha~H$\alpha$~10 &      M6.25 & 12.79$\pm$0.01 & 12.68$\pm$0.02 & 12.61$\pm$0.05 & 12.59$\pm$0.06   \\
    ISO~138 &       M6.5 & 12.50$\pm$0.02 & 12.21$\pm$0.01 & 11.97$\pm$0.03 & 11.30$\pm$0.02   \\
 2MASS~11082570-7716396 &         M8 &    \nodata & 13.01$\pm$0.01 &    \nodata & 11.95$\pm$0.03   \\
 2MASS~11093543-7731390 &      M8.25 & 13.81$\pm$0.01 & 13.70$\pm$0.01 & 13.63$\pm$0.05 & 13.53$\pm$0.04   \\
   Cha~H$\alpha$~1  &      M7.75 & 11.56$\pm$0.01 & 11.15$\pm$0.02 & 10.76$\pm$0.01 &  9.67$\pm$0.02   \\
   Cha~H$\alpha$~7  &      M7.75 & 11.89$\pm$0.01 & 11.75$\pm$0.01 & 11.70$\pm$0.04 & 11.63$\pm$0.03   \\
\enddata
\tablenotetext{a}{From \citet{luh04cha}, \citet{luh04bin}, \citet{luh04ots}, 
and Luhman (in preparation).}
\end{deluxetable}


\end{document}